\documentclass[12pt]{article}
\usepackage{amssymb}
\usepackage{a4}
\def\b#1{{\mathbb #1}}
\def\c#1{{\cal #1}}

\def\nn{\nonumber \\}

\def\1{{\bf 1}}
\def\0{{\bf 0}}

\def\RH{\mbox{$\hat {\sf R}$\,}}

\def\P{\mbox{$\cal P\!\!$}}

\def\x{\mbox{x}}

\def\A{\mbox{$\cal A$}}
\def\id{\mbox{id\,}}

\newcommand{\uot}{\underline{\otimes}}
\newcommand{\und}{\underline}

\newcommand{\be}{\begin{equation}}
\newcommand{\ee}{\end{equation}}
\newcommand{\bea}{\begin{eqnarray}}
\newcommand{\eea}{\end{eqnarray}}
\newcommand{\ba}{\begin{array}}
\newcommand{\ea}{\end{array}}
\def\sq{\mbox{\rlap{$\sqcap$}$\sqcup$}}
\newenvironment{proof}[1]{\vspace{5pt}\noindent{\bf Proof #1}\hspace{6pt}}
{\hfill\sq}
\newcommand{\bp}{\begin{proof}}
\newcommand{\ep}{\end{proof}\par\vspace{10pt}\noindent}
\begin{document}

\title{$q$-Deformed  $su(2)$ instantons by $q$-quaternions
}

\author{        Gaetano Fiore, \\\\
         \and
        Dip. di Matematica e Applicazioni, Universit\`a ``Federico II''\\
        V. Claudio 21, 80125 Napoli, Italy\\
        \and
        \and
        I.N.F.N., Sezione di Napoli,\\
        Complesso MSA, V. Cintia, 80126 Napoli, Italy
        }
\date{}

\maketitle \abstract{Interpreting the coordinates of the quantum
Euclidean space $\b{R}_q^4$ [the $SO_q(4)$-covariant noncommutative
space] as the entries of a ``$q$-quaternion matrix'' we construct
(anti)instanton solutions of a would-be $q$-deformed $su(2)$
Yang-Mills theory on $\b{R}_q^4$. Since the (anti)selfduality
equations are covariant under the quantum group of deformed
rotations, translations and scale change, by applying the latter we
can respectively generate ``gauge equivalent'' or ``inequivalent''
solutions from the  one centered at the origin and with unit size.
We also construct multi-instanton solutions. As these solutions
depend on noncommuting parameters playing the roles of `sizes' and
`coordinates of the centers' of the instantons, this indicates that
the moduli space of a complete theory should be a noncommutative
manifold. Similarly, as the (global) gauge transformations relating
``gauge equivalent'' solutions depend on the generators of two
copies of $SU_q(2)$, this suggests that gauge transformations should
be allowed to depend on additional noncommutative parameters. }

\vfill
\noindent
- Preprint 07-04 Dip. Matematica e Applicazioni, Universit\`a di Napoli;\\
\noindent
- DSF/4-2007
\newpage

\section{Introduction}

A broad attention has been devoted in recent years to
the construction of gauge field theories on noncommutative manifolds.
A crucial test
of this construction is the search of instantonic solutions, especially
after the discovery \cite{NekSch98}  that deforming $\b{R}^4$ into
the Moyal-Weyl noncommutative Euclidean space $\b{R}_{\theta}^4$ regularizes
the zero-size singularities of the instanton moduli space (see also
\cite{SeiWit99}).
Various other noncommutative geometries have been considered
(see e.g. \cite{ConLan01,BonCicTar00,DabLanMas01,LanvSu06}). They do not
always completely fit Connes' standard framework of noncommutative geometry
\cite{Con94}, thus stimulating attempts of generalizations. Among the available
deformations of $\b{R}^4$ there is also the Faddeev-Reshetikhin-Takhtadjan
noncommutative Euclidean space $\b{R}_q^4$ covariant under $SO_q(4)$
\cite{FadResTak89}. This, as other quantum group covariant
noncommutative spaces (shortly: quantum spaces), is maybe even more
problematic for the formulation \cite{JurMoeSchSchWes01} of a gauge field
theory. One main reason is the lack of a proper (i.e.
cyclic) trace to define gauge invariant observables (action, etc). Another one
is the complicated $\star$-structure of the differential calculus
for real $q$. Here, leaving these
two issues aside, we formulate and solve the (anti)selfduality equations on it;
we omit mathematical details and proofs, which can be found in the longer paper
\cite{Fio06}.
This  might contribute to suggest more general
formulations of gauge theories on noncommutative manifolds (include
quantum spaces) where e.g. gauge transformations, gauge potentials, and the corresponding field
strengths depend not only on coordinates, but also on derivatives
(as suggested e.g. in \cite{DimMeyMoeWes03,AscDimMeySchWes06}) and/or
possibly on additional noncommuting parameters (see section \ref{shift} below).

As known, the search and classification \cite{AtiDriHitMan78} of
Yang-Mills instantons on $\b{R}^4$ is largely simplified
when the latter is promoted to the
quaternion algebra $\b{H}$.
Following the undeformed case, we introduce (section \ref{quater})
a notion of a $q$-quaternion as the
defining matrix of a copy of $\b{H}_q:=SU_q(2)\times
\b{R}^{\ge}$ ($\b{R}^{\ge}$ denoting the semigroup of nonnegative
real numbers)  and reformulate
the algebra $\A$ of functions on $\b{R}_q^4$ as a $\star$-bialgebra
$C\big(\b{H}_q\big)$.  The bialgebra structure encodes the property that the
product of two quaternions is a quaternion and is inherited from the bialgebra of
$2\times 2$ quantum matrices \cite{Dri86,FadTak86,Wor87a,FadResTak89}
(therefore it differs from the proposal in \cite{Maj94}).
It also turns out that the quantum sphere $S_q^4$
of \cite{DabLanMas01}
can be regarded as a compactification of this $\star$-algebra.
In section \ref{covdiffcal} we reformulate in $q$-quaternion language  the
$SO_q(4)$-covariant differential calculus, the $SO_q(4)$-covariant $q$-epsilon tensor and Hodge
map \cite{Fio94,Fio04JPA} on $\Omega^*(\b{R}^4_q)$.  In section \ref{instconst} we formulate (anti)selfduality
equations and find a large family of solutions $A$
in the form of 1-form valued $2\times 2$
matrices both in the ``regular'' and in the ``singular gauge''. There is a
larger indeterminacy than in the undeformed theory because we are not yet able
to formulate and impose the correct antihermiticity condition on the gauge
potential. Among the solutions there are some distinguished choices that
 closely resemble (in $q$-quaternion language) their undeformed counterparts (instantons and anti-instantons) in
$su(2)$ Yang-Mills theory on $\b{R}^4$. [The (still missing)
complete gauge theory might however be  a deformed $u(2)$ rather
than $su(2)$ Yang-Mills theory.] The projector characterizing the instanton
projective module (playing the role of the vector bundle) of \cite{DabLanMas01}
in $q$-quaternion language takes exactly the same natural form as in
the undeformed theory.
We also point out where the present model  doesn't fit
the today standard formulation \cite{Con94} of gauge
theory on noncommutative spaces (some basic notions of
which we recall in section \ref{gaugeth}). In analogy with the
undeformed (and the Nekrasov-Schwarz \cite{NekSch98}) case,
applying (section \ref{shift}) the quantum group $SO_q(4)$
of $q$-deformed rotations  one obtains  gauge
equivalent solutions (by a global gauge transformation),
whereas applying $q$-deformed dilatations and (the braided group
of) $q$-deformed translations one finds
gauge inequivalent solutions; however
this global gauge transformation depends on new
noncommuting parameters playing the role of coordinates of $SO_q(4)$,
and the gauge inequivalent solutions
depend on the noncommuting ``coordinates of the center''
of the (anti)instanton.
Finally (section \ref{multiinst}), we find  first $n$-instantons solutions
 in the ``singular'' gauge for any integer $n$; the construction procedure
is not yet the deformed analog of the general ADHM one \cite{AtiDriHitMan78},
but rather of the procedure initiated in \cite{tHo76} and developed
in \cite{vari}, which reduces to the determination of a suitable harmonic scalar
potential, expressed in quaternion language. Then for $n=1,2$ we transform
the singular solutions into ``regular'' solutions by ``singular gauge
transformations'', as in the undeformed case (of course the $n=1$ regular
instanton solution is again one found in section \ref{instconst}). The solutions
are parametrized by noncommuting parameters playing the role of ``sizes'' and
``coordinates of the centers'' of the (anti)instantons. This indicates that
the moduli space of a complete theory will be a noncommutative manifold.
This is  similar to what was proposed
in \cite{IvaLecMue04} for $\b{R}_{\theta}^4$ for selfdual
deformation parameters $\theta_{\mu\nu}$.

\section{The $q$-quaternion
bialgebra $C\left(\b{H}_q\right)$}
\label{quater}

We start by recalling how the (undeformed) quaternion
$\star$-algebra $\b{H}$
can be formulated in terms of $2\times 2$ matrices:
any $X\in \b{H}$ is given by
\[
X=\x_1+\x_2i+\x_3j+\x_4k,
\]
with $\x\in \b{R}^4$ and imaginary $i,j,k$ fulfilling
\[
i^2=j^2=k^2=-1,\qquad\qquad  ijk=-1.
\]
One refers to $\x_1$ and to the following three terms as to the `real'
and `imaginary' part of $X$ respectively.
Replacing $i,j,k$ by Pauli matrices times the imaginary unit ${\rm
i}$ we can associate to $X$ a matrix
\[ X\leftrightarrow x\equiv\left(\ba{cc}
\x_1+\x_4{\rm i} \:& \x_3+\x_2{\rm i}\\
-\x_3+\x_2{\rm i} \:& \x_1-\x_4{\rm i} \ea\right)=: \left(\ba{cc}
\alpha\:& -\gamma^{\star}\\
\gamma\:& \alpha^{\star} \ea\right) \] (where $\alpha,
\gamma\in\b{C}$).
The quaternionic product becomes represented
by matrix multiplication, and the  quaternionic
conjugation becomes represented
by hermitean conjugation of the matrix $x$.
  Therefore  $\b{H}$ can be seen also
as the subalgebra of $M_2(\b{C})$ consisting
of all complex $2\times 2$ matrices of this
form. Since the determinant of any $x$ is nonnegative,
$$
|x|^2\equiv\det(x)=|a|^2+|\gamma|^2 \ge 0,
$$
any $x$ can be factorized in the form
$ x=T|x|$,
where $T\in SU(2)$ and $|x|$ belongs
to the semigroup $\b{R}^{\ge}$ of nonnegative real numbers.
Hence any $x$ belongs also to the
semigroup  $SU(2)\times \b{R}^{\ge}$.

\medskip
We $q$-deform this just replacing $SU(2)$
by $SU_q(2)$ in the algebra
of functions of the matrix elements of $x$. In other words, we define a
$q$-quaternion just as one introduces the defining matrix of
$SU_q(2)$ \cite{Wor87,Wor87a}, but without imposing the
unit $q$-determinant condition.
For $q\in\b{R}$ consider the unital associative
$\star$-algebra $\A\equiv C(\b{H}_q)$ generated by elements $\alpha,
\gamma^{\star},\alpha^{\star},\gamma$ fulfilling the commutation
relations
\be
\ba{l}
\alpha\gamma=q\gamma\alpha,\qquad\alpha\gamma^{\star}=q\gamma^{\star}\alpha,
\qquad\gamma\alpha^{\star}=q\alpha^{\star}\gamma,\\[8pt]
\gamma^{\star}\alpha^{\star}=q\alpha^{\star}\gamma^{\star},\qquad
[\alpha,\alpha^{\star}]=(1\!-\!q^2)\gamma\gamma^{\star} \qquad
[\gamma^{\star},\gamma]=0.\ea \label{explqquatcomrel}
\ee
Introducing the matrix
\be
x\equiv \left(\ba{ll} x^{11}\: & x^{12}\\
x^{21}\:& x^{22}\ea\right):=\left(\ba{cc} \alpha\:& -q\gamma^{\star}\\
\gamma\: & \alpha^{\star} \ea\right)
\ee
we can rewrite these commutation relations as
\be
\hat R x_1x_2=x_1x_2\hat R      \label{qquatcomrel}
\ee and the conjugation relations as
$x^{\alpha\beta}{}^{\star}=\epsilon^{\beta\gamma}x^{\delta\gamma}
\epsilon_{\delta\alpha}$, i.e.
\be
x^{\dagger}=\bar x\qquad\qquad
\mbox{where } \bar a:=\epsilon a^T\epsilon^{-1} \quad\forall a\in
M_2. \label{qquatstarrel}
\ee
Here as usual $x_1\equiv
x\otimes_{\b{C}} I_2$, $x_2\equiv I_2\otimes_{\b{C}} x$ ($I_2$ is
the $2\times 2$ unit matrix), $\hat R$ is the braid matrix of
$M_q(2)$, $GL_q(2)$ and $SU_q(2)$ \be \hat
R^{\alpha\beta}_{\gamma\delta}=q\delta^{\alpha}_{\gamma}
\delta^{\beta}_{\delta}+\epsilon^{\alpha\beta}\epsilon_{\gamma\delta},
                                               \label{Rexpl}
\ee
and $\epsilon$ is the corresponding completely
antisymmetric tensor
\be
\epsilon\equiv (\epsilon_{\alpha\beta})
:=\left(\ba{cc} 0\: & 1\\ -q \: & 0
\ea\right),
\qquad\qquad
\epsilon^{-1}\equiv (\epsilon^{\alpha\beta})
=-q^{-1}(\epsilon_{\alpha\beta}).      \label{qepsilon}
\ee
The decomposition of $\hat R$
in orthogonal projectors reads \be \hat
R=q\P_s-q^{-1}\P_a, \ee  and the $q$-deformed symmetric,
antisymmetric projectors
$\P_s,\P_a$  can be expressed as
\be
\c{P}_a{}^{\alpha\beta}_{\gamma\delta}=-
\frac{\epsilon^{\alpha\beta}\epsilon_{\gamma\delta}}{q+q^{-1}},
\qquad\qquad
\c{P}_s{}^{\alpha\beta}_{\gamma\delta}=\delta^{\alpha}_{\gamma}
\delta_{\delta}^{\beta}+
\frac{\epsilon^{\alpha\beta}\epsilon_{\gamma\delta}}{q+q^{-1}}.
\label{exproj}
\ee
$\A:=C(\b{H}_q)$ can be naturally endowed with
a $\star$-bialgebra  structure
(we are not excluding ${\bf 0}_2$ from the spectrum of $x$),
more precisely the above real section
 of the bialgebra $C\left(M_q(2)\right)$ of
$2\times 2$ quantum matrices \cite{Dri86,FadTak86,Wor87a,FadResTak89}.
In the sequel we shall write the corresponding coproduct
$\Delta(x^{\alpha\gamma})\!=\! x^{\alpha\beta}\!\otimes\! x^{\beta\gamma}$
in the more compact matrix product form
\be
x\to \Delta(x)=ax                                 \label{coprod}
\ee
where we have renamed $x\otimes \1\to a$, $\1\otimes x\to x$.
Since  the coproduct
is a $\star$-algebra map, $\Delta(x)$, or equivalently the matrix product $ax$ of any
two matrices $a,x$ with mutually commuting entries and fulfilling
(\ref{qquatcomrel}-\ref{qquatstarrel}), again fulfills the latter.
Therefore we shall call any such matrix $x$ a {\it $q$-quaternion},
and $\A:=C(\b{H}_q)$ the $q$-quaternion bialgebra. Note that, according
to this definition, $I_2$ is a $q$-quaternion, and $x$ is a $q$-quaternion
iff $-x$ is.
As well-known, the socalled
`$q$-determinant' of $x$ \be |x|^2
\equiv\det{}_q(x):=x^{11}x^{22}\!-\!qx^{12}x^{21}=\alpha^{\star}\alpha
+ \gamma^{\star}\gamma = \frac 1{1\!+\!q^2}
x^{\alpha\alpha'}\! x^{\beta\beta'}\!
\epsilon_{\alpha\beta}\epsilon_{\alpha'\beta'}, \label{qdet} \ee
is
central, manifestly nonnegative-definite and group-like. Therefore in
any $\star$-representation it will have zero eigenvalue
iff $x$ has $\0_2$ as an eigenvalue matrix. Replacing (\ref{Rexpl}) in
(\ref{qquatcomrel}) we find that the latter is equivalent  to
\be
x\bar x=\bar xx=|x|^2I_2.                \label{blu}
\ee

If we extend $\A=C(\b{H}_q)$ also by the new (central,
positive-definite and group-like) generator $|x|^{-1}$ (this will
exclude $x=\0_2$ from the spectrum),  the matrix $x$ becomes
invertible and we obtain even a Hopf $\star$-algebra  with antipode
$S$ defined by \be Sx=x^{-1}=\frac{\bar x}{|x|^2}, \qquad\qquad S
|x|^{-1}=|x|. \label{inverse} \ee The matrix elements of $T:=\frac
x{|x|}$ fulfill the `RTT' \cite{FadResTak89} relations
(\ref{qquatcomrel}) and \be T^{\dagger}=T^{-1}=\overline{T},
\qquad\qquad\det{}_q(T)=\1, \label{Tinverse} \ee namely generate
$C\left(SU_q(2)\right)$ \cite{Wor87,Wor87a} as a quotient
subalgebra. Therefore in this case the $x^{\alpha\alpha'}$  generate
the (Hopf) $\star$-algebra $C\left(SU_q(2)\!\times\! GL^+(1)\right)$
of functions on the ``quantum group $SU_q(2)\times GL^+(1)$ of
non-vanishing $q$-quaternions'' [a real section of the Hopf algebra
$C\left(GL^+_q(2)\right)$], in analogy with the $q=1$ case.

\bigskip
One can easily verify that as a $\star$-algebra $\A:=C(\b{H}_q)$ coincides
with the algebra of functions on the $SO_q(4)$-covariant quantum Euclidean
Space $\b{R}_q^4$ of \cite{FadResTak89}. We identify the present
$qx^{11},x^{12},-qx^{21},x^{22}$ with the
generators $x^1,x^2$, $x^3,x^4$ of \cite{FadResTak89} (in their
original indices convention)
or with  the generators $x^{-2},x^{-1}$, $x^1,x^2$ in
 the convention of Ref.  \cite{Ogi92} (which has been
heavily used by the author).

\medskip
The algebra and the $\star$-structure are covariant under, i.e.
preserved by, matrix multiplication
$$
x\to a\, x\,b
$$
by the defining matrices
$a,b$ of two copies
$SU_q(2)$, $SU_q(2)'$ of the special unitary quantum
group, or of two copies $\b{H}_q$, $\b{H}_q'$ of the
quaternion quantum group, if the entries of $a,b$
commute with each other
and with the entries of $x$. In other words they are covariant
under the (mixed left-right) coactions of $SU_q(2)\otimes SU_q(2)'=Spin_q(4)$ and
of $\b{H}_q\otimes \b{H}_q'$. This follows from the fact that the twofold coproduct
$\Delta^{(2)}(x)=axb$,  \be
\Delta^{(2)}(x^{\alpha\alpha'})=a^{\alpha\beta}b^{\beta'\alpha'}\otimes
x^{\beta\beta'},\qquad \qquad \mbox{i.e.}\quad
x\stackrel{\Delta^{(2)}}{\longrightarrow} a\, x\,b,  \label{SUq2SUq2coaction}
\ee
is a  $\star$-homomorphism, or equivalently both the
the left coaction $x\to a\, x$ and the right one $x\to x\,b$ are.
In terms of $x^i$ this takes the form
\be
\Delta^{(2)}(x^i)={\bf T}^i_j\otimes
x^j,\qquad \qquad  {\bf T}^i_j:= B^i_{\alpha\alpha'}
a^{\alpha\beta}b^{\beta'\alpha'}B^{-1}{}^{\beta\beta'}_j,
\label{SOq4coaction}
\ee
where
$B\equiv(B^a_{\alpha\alpha'})$  is the (diagonal and invertible) matrix entering
the linear transformation $x^a=B^a_{\alpha\alpha'}x^{\alpha\alpha'}$.
Relation (\ref{SOq4coaction})$_1$
 has the same form as the left coaction of Ref. \cite{FadResTak89}
of the quantum group $SO_q(4)$ [and of its
extension $\widetilde{SO_q(4)}:=SO_q(4)\!\times\! GL^+(1)$,
the quantum group of rotations and scale transformations in 4
dimensions] on $\b{R}_q^4$. This is no formal coincidence: the
${\bf T}^i_j$ fulfill the $RTT$ commutation relations and $\star$-conjugation relations
\be
\RH{\bf T}_1{\bf T}_2={\bf T}_1{\bf T}_2\RH, \qquad\qquad
{\bf T}^i_j{}^{\star}=g^{jj'}{\bf T}^{i'}_{j'}g_{i'i}  \label{defSOq4}
\ee
and in addition $g_{ii'}{\bf T}^i_j{\bf T}^{i'}_{j'}=g_{jj'}\1$ if the central
element $|a||b|$ is 1 (here $\RH$ and $g_{ab}=B^{-1}{}^{\alpha\alpha'}_aB^{-1}{}^{\beta\beta'}_b\epsilon_{\alpha\beta}\epsilon_{\alpha'\beta'}$ are the
braid matrix and the metric matrix of $SO_q(4)$).   These are respectively
the  defining relations
of  $\widetilde{SO_q(4)}$ and of  the compact quantum subgroup $SO_q(4)$
\cite{FadResTak89}. We have thus an explicit realization
of the equivalences
$$
SO_q(4)=SU_q(2)\!\times\! SU_q(2)'/\b{Z}_2 ,
\qquad\quad\widetilde{SO_q(4)}=\b{H}_q\!\times\!\b{H}_q'/GL(1).
$$
The quotient over $\b{Z}_2$ is due to the invariance of
${\bf T}^i_j $ under $(a,b)\to (-a,-b)$.

As we shall recall in section \ref{shift}, the commutation relations are also
invariant  under the braided group of translations \cite{Maj92,Maj95} $\b{R}_q^4$,
which is the $q$-deformed version of the group of translations
$\b{R}^4$; the
role of composition of translations is played by the socalled braided
coaddition.
They are in fact covariant under the coaction of the full
inhomogenous extension $\widetilde{ISO_q(4)}$ \cite{SchWeiWei92} of
$\widetilde{SO_q(4)}$ (or quantum Euclidean group in 4 dimensions), which
includes $q$-deformed translations together with scale changes and rotations
($\widetilde{ISO_q(4)}$ can be obtained also by ``bosonization''
of $\b{R}_q^4$ \cite{Maj92}).

\subsubsection*{Comparison and links with other formulations}

A matrix version of the 4-dim quantum Euclidean
space (with no interpretation in terms of $q$-deformed quaternions)
was proposed also in \cite{Maj94}. However, the
$\star$-relations and the $SO_q(4)$-coaction are different, i.e. cannot be put
both in the form (\ref{explqquatcomrel}), (\ref{SUq2SUq2coaction}), even by a
relabelling of the generators.

\medskip
The slightly extended $\star$-algebra  $\A^{ext}$ obtained
by adding as generators the central elements
$1/\left(1+\frac{|x|^2}{\rho^2}\right)$, $\rho\in\b{R}^+$,
contains the $\star$-algebra of functions on the quantum 4-sphere
$S_q^4$ proposed in \cite{DabLanMas01} (as a
`suspension' of  the algebra of a quantum 3-sphere $S_q^3$). Define
\be
\alpha'=\alpha^{\star} \frac {2\sqrt{2}}{1\!+\!2|x|^2}e^{ia},\qquad\quad
\beta'=\gamma^{\star} \frac {2\sqrt{2}}{1\!+\!2|x|^2}e^{ib},\qquad\quad
 z= \frac {1\!-\!2|x|^2}{1\!+\!2|x|^2},
\label{redef}
\ee
where $\alpha,\gamma,\alpha^{\star},\gamma^{\star}$ fulfill
(\ref{explqquatcomrel})
and $e^{ia},e^{ib}\in U(1)$ are possible
phase factors.
Then $\alpha',\beta',z$ fulfill the defining relation (1) of the
$C^{\star}$-algebra considered in Ref. \cite{DabLanMas01}
(where these elements are respectively denoted as $\alpha,\beta,z$),
in particular
\be
\alpha'\alpha'{}^{\star}+\beta'\beta'{}^{\star}+z^2=\1,
\ee
and the invertible function $z(|x|)$ spans
$[-1,1[$, i.e. all the spectrum of $z$ except the eigenvalue $z=1$,
as $|x|$ spans all its spectrum $[0,\infty[$.
Viceversa, starting from the latter and enlarging it so that
it contains the element $(1\!+\!z)/2(1\!-\!z)=:|x|^2$
then inverting the above formulae
one obtains elements $\alpha,\gamma,\alpha^{\star}\gamma^{\star}$ fulfilling
our defining  relations (\ref{explqquatcomrel}).

The redefinitions
(\ref{redef}) have exactly the form of a stereographic
projection of $\b{R}^4$ on a sphere $S^4$ of unit radius
(the square radius is $x\cdot x=2|x|^2$): $S^4$ is the sphere centered at the
origin and $\b{R}^4$ the subspace $z=0$ immersing both in a $\b{R}^5$ with
coordinates defined by $X\equiv(Re(\alpha'),Im(\alpha'),
Re(\beta'),Im(\beta'),z)$.
In the commutative theory adjoining the missing point $X=(0,0,0,0,1)$ of
$S^4$ amounts to adding to $\b{R}^4$ the point at infinity, i.e.
to compactifying $\b{R}^4$  to $S^4$. We can thus regard
the transition from our algebra to the one  considered in Ref.
\cite{DabLanMas01}
as a compactification of $\b{R}_q^4$ into their $S^4_q$.

\section{Other preliminaries}
\label{covdiffcal}

The $SO_q(4)$-covariant differential calculus \cite{CarSchWat91}
$(d,\Omega^*)$
on $\b{R}_q^4\sim\b{H}_q$ is obtained imposing covariant
homogeneous bilinear commutation relations
 (\ref{xxirel}) between the $x^a$ and the differentials
$\xi^b:=dx^b$.
 Partial derivatives are introduced through the
decomposition
$d=\xi^a\partial_a=\xi^{\alpha\alpha'}\partial_{\alpha\alpha'}$ of
the ($SO_q(4)$-invariant) exterior derivative. All other commutation
relations are derived by consistency with nilpotence and the Leibniz
rule. Beside (\ref{qquatcomrel}),  we have \bea &&
x^{\alpha\alpha'}\xi^{\beta\beta'}= \hat
R^{\alpha\beta}_{\gamma\delta} \hat
R^{\alpha'\beta'}_{\gamma'\delta'}
\xi^{\gamma\gamma'}x^{\delta\delta'},\label{xxirel}\\
&& \P_s{}^{\alpha\beta}_{\gamma\delta}
\P_s{}^{\alpha'\beta'}_{\gamma'\delta'}\xi^{\gamma\gamma'}
\!\xi^{\delta\delta'}\!=\!0\!=\!(\xi\epsilon\xi^T)^{\gamma\delta}
\epsilon_{\gamma\delta},\qquad\quad\label{xixirel}\\ &&
\partial_{\alpha\alpha'}\partial_{\beta\beta'}= \hat
R^{\delta\gamma}_{\beta\alpha} \hat
R^{-1}{}^{\delta'\gamma'}_{\beta'\alpha'}
\partial_{\gamma\gamma'}\partial_{\delta\delta'}, \label{ddrel}\\
&& \partial_{\alpha\alpha'}
x^{\beta\beta'}\!\!=\!\delta^{\beta}_{\alpha}
\delta^{\beta'}_{\alpha'}\!+\! \hat R^{\beta\delta}_{\alpha\gamma}
\hat R^{\beta'\delta'}_{\alpha'\gamma'}
x^{\gamma\gamma'} \!\partial_{\delta\delta'}, \quad\:    \label{dxrel}\\
&& \partial_{\alpha\alpha'} \xi^{\beta\beta'}= \hat
R^{-1}{}^{\beta\delta}_{\alpha\gamma} \hat
R^{-1}{}^{\beta'\delta'}_{\alpha'\gamma'}\xi^{\gamma\gamma'}
\partial_{\delta\delta'}.\label{dxirel}
\eea [An alternative $SO_q(4)$-covariant differential calculus
$(\hat d,\hat\Omega^*)$ is obtained replacing $\hat R$ by $\hat
R^{-1}$ in (\ref{xxirel}-\ref{dxirel})]. The $\xi^i$ transform under
$SO_q(4)$ exactly as the $x^i$, the $\partial_i$ in the
contragradient corepresentation. In terms of $x^i,\partial_j$ one can build a
special unitary element $\lambda$ such that \be \lambda
x^i=q^{-1}x^i\lambda,\qquad\quad
\lambda\partial^i=q\partial^i\lambda, \qquad\quad \lambda
\xi^i=\xi^i\lambda.                        \label{Lambdaprop} \ee We
introduce the notation $
\partial^{\alpha\alpha'}\!:=\!\epsilon^{\alpha\beta}
\epsilon^{\alpha'\beta'}\partial_{\beta\beta'}$, $
\partial\equiv \big(\partial^{\alpha\alpha'}\big)$.  The
$\partial^{\alpha\alpha'}$ fulfill the same commutation relations
(among themselves) as the $x^{\alpha\alpha'}$, and transform in the
same way under the $SO_q(4)$ coaction. As a consequence, the
Laplacian $\Box:=g^{hk}\partial_k\partial_h=
\partial^{\alpha\alpha'}\partial_{\alpha\alpha'}$
is $SO_q(4)$-invariant and commutes with the
$\partial_{\beta\beta'}$, and \be
\partial\bar\partial=\bar\partial\partial=
I_2|\partial|^2\equiv I_2\frac 1{1\!+\!q^2}\Box . \ee From
(\ref{dxrel}), (\ref{dxirel}) it follows \be
q^2\partial|x|^2\!=\!x\!+\!q^4|x|^2\partial,\qquad 
\partial\frac 1{|x|^2}\!=\!-\frac{q^{-4}x}{|x|^4}
\!+\!\frac {q^{-2}}{|x|^2}\partial,\qquad 
|\partial|^2 \frac 1{|x|^2}\!=\!  \frac {q^{-4}}{|x|^2}
|\partial|^2\!-\! \frac {q^{-6}}{|x|^4} x\!\cdot\! \partial\label{utile4}
\ee
Since the rhs of the latter formula applied to $\1$ gives zero, $1/|x|^2$ is
harmonic, as in the undeformed case.

\medskip
We denote as  ${\cal DC}^*$
(``differential calculus algebra'') the algebra (over $\b{C}$)
generated by $\1, x^i,\xi^i,\partial_i,\lambda^{\pm 1}$; the elements are
differential-operator-valued forms.
We also denote as $\tilde\Omega^*$ the unital subalgebra generated by
$\xi^i,x^i,\lambda^{\pm 1}$ ,  as $\Omega^*$ (algebra of differential forms)
the unital subalgebra generated by $\xi^i,x^i$, as $\Omega_S^*$ the unital
subalgebra  generated by
$T^{\alpha\alpha'},dT^{\alpha\alpha'}$,  as  $\bigwedge^*$
(algebra of exterior forms) the unital subalgebra generated
by $\xi^i$.

As usual we introduce in these algebras a grading
$\natural\in\b{N}$ given by the degree in $\xi^i$, and denote as
${\cal DC}^p,\Omega^p$, etc., their components with $\natural =p$.
Each of these  components  is a bimodule of dimension $4\choose{p}$
w.r.t. to its 0-component. For instance, since
by definition $\Omega^0=\A$, $\Omega^p$ is a $4\choose{p}$-dimensional
$\A$-bimodule; similarly, ${\cal DC}^p$ is a $4\choose{p}$-dimensional
${\cal H}$-bimodule, where ${\cal H}:={\cal DC}^0$ (the Heisenberg algebra),
generated by the $x^i, \partial_i,\lambda^{\pm 1}$.
Any $\bigwedge^p$ carries an irreducible corepresentation of $SO_q(4)$.
In particular, as ${4\choose{4}}=1$ all exterior 4-forms
are $SO_q(4)$-invariant and proportional to
$d^4x:=\xi^1\xi^2\xi^3\xi^4$.

All this is exactly as in the case $q\!=\!1$, except that as a
$C\big(SU_q(2))$-bimodule
$\Omega_S^p$ is $4\choose{p}$-dimensional
when $q\!\neq\! 1$ and $3\choose{p}$-dimensional when $q\!=\!1$.

\medskip
The whole set of commutation relations (\ref{qquatcomrel}),
(\ref{xxirel}-\ref{dxirel})  is  \cite{CerFioMad01}  invariant under
the  replacement
$x^{\alpha\alpha'}/|x|^2q^2(1\!-\!q^2)\to\partial^{\alpha\alpha'}$.
As a corollary, on $\Omega^*$ one can realize the action of the
exterior derivative as the (graded) commutator \be
d\omega_p=[-\theta,\omega_p\}:=-\theta\omega_p+(-)^p\omega_p\theta,
\qquad\qquad \omega_p\in\Omega^p \label{thetacommu} \ee with the
special $SO_q(4)$-invariant 1-form \cite{Ste96} (the `Dirac
Operator', in Connes' \cite{Con94} parlance) \be
\theta:=(d|x|^2)|x|^{-2}\frac 1{q^2-1}=
\frac{q^{-2}}{q^2-1}\xi^{\alpha\alpha'}\frac{x^{\beta\beta'}}{|x|^2}
\epsilon_{\alpha\beta}\epsilon_{\alpha'\beta'}. \label{deftheta} \ee
$\theta$ is closed and singular in the $q\to 1$ limit. Applying $d$ to (\ref{blu}) we find \be
x\bar\xi+\xi\bar x=(q^2\!-\!1)\theta|x|^2I_2,\qquad\qquad \bar
x\xi+\bar\xi x=(q^2\!-\!1)\theta|x|^2 I_2.   \label{xblu} \ee
Relation (\ref{xxirel}) implies $|x|^2\xi^i=q^2\xi^i|x|^2$, which we
supplement with the compatible ones \be
q|x|^{-1}\xi^i=\xi^i|x|^{-1}, \qquad\Rightarrow\qquad
q\,|x|^{-1}\,\theta=\theta\,|x|^{-1}. \label{theta|x|rel} \ee By a
straightforward computation one also finds \be
 dT^{\alpha\alpha'}=
q^{-1}\xi^{\alpha\alpha'}\frac 1{|x|}+(q^{-1}\!-\!1)\theta T^{\alpha\alpha'}.
\label{dT}
\ee
By (\ref{SUq2SUq2coaction}) the 1-form-valued $2\times 2$ matrices
$(dT)\overline{T}$, $(d\overline{T})T$ are manifestly invariant
under respectively the right and left coaction of $SU_q(2)$, or
equivalently the $SU_q(2)'$ and  the $SU_q(2)$ part of $SO_q(4)$
coaction. Setting $Q:=-\epsilon^{-1}\epsilon^T$ one finds
$$
\mbox{tr}[Q(dT)\overline{T}]=\mbox{tr}[Q^{-1}(d\overline{T})T]=(q\!-\!1)(q\!-\!q^{-2})
\theta;
$$
from (\ref{deftheta}) we see that
only in the $q\to 1$ limit these traces vanish. That's why for generic
$q\neq 1$ the four matrix elements of either $(dT)\overline{T}$ or
$(d\overline{T})T$ are independent, and make up alternative bases
for both $\Omega^*_S$ and $\Omega^*$.

Actually, one can check (we will give details in \cite{Fio06corfu})
that $(d,\Omega^*)$ coincides
with the bicovariant differential calculus on $M_q(2),GL_q(2)$
\cite{S2,SchWatZum92}, and
$(d,\Omega^*_S)$ coincides
with the Woronowicz 4D-  bicovariant one \cite{Wor89,PodWor90} on $C\big(SU_q(2)\big)$.

\bigskip
One major problem in the present $q\in\b{R}$ case is that the calculus is not
real: there is no $\star$-structure such that
$d(f^{\star})= (df)^{\star}$, nor is there a $\star$-structure
$\star:\Omega^*\to\Omega^*$.
Formally, a $\star$-structure would map the commutation relations of
$(d,\Omega^*)$ into the ones of $(\hat d,\hat\Omega^*)$, and conversely.
At least, there is a $\star$-structure  \cite{OgiZum92}
$$
\star:{\cal DC}^*\to {\cal DC}^*
$$
having the desired commutative limit (the  $\star$-structure of the De Rham
calculus on $\b{R}^4$), but a rather nonlinear character (incidentally, the
latter  has been recently \cite{Fio04} recast in a much more suggestive form),
in other words objects of the second calculus can be realized nonlinearly in
terms of objects of the first (and conversely).

\bigskip
The {\bf Hodge map}  is a $SO_q(4)$-covariant,
$\A$-bilinear map $*:\tilde\Omega^p\to\tilde\Omega^{4-p}$
\cite{Fio04JPA}  such that
$*^2= \id$, defined by
\[
{}^*(\xi^{i_1}...\xi^{i_p})= c_p\,\xi^{i_{p+1}}...\xi^{i_4}
\varepsilon_{i_4...i_{p+1}}{}^{i_1...i_p}\lambda^{2p-4},
\]
where $\varepsilon^{hijk}\equiv$ $q$-epsilon tensor \cite{Fio04JPA,Fio94}
and $c_p$ are suitable normalization factors.
Actually this extends \cite{Fio04JPA} to a ${\cal H}$-bilinear map
$*:{\cal DC}^p\to{\cal DC}^{4-p}$ with the same features. For $p=2$
$\lambda$-powers disappear and one even gets maps
$*:\Omega^2\to\Omega^2$, $*:{\cal DC}^2\to{\cal DC}^2$. The previous
equation becomes \be {}^*f^{\alpha\beta}=f^{\alpha\beta}\qquad\qquad
{}^*f'{}^{\alpha'\beta'}=-f'{}^{\alpha'\beta'} \ee in terms of the
``selfdual exterior 2-forms'' \be
f^{\alpha\beta}:=\P_s{}^{\alpha\beta}_{\gamma\delta}
\epsilon_{\gamma'\delta'}\xi^{\gamma\gamma'}\xi^{\delta\delta'}=
\epsilon_{\gamma'\delta'}\xi^{\alpha\gamma'}\xi^{\beta\delta'}
=(\xi\epsilon\xi^T)^{\alpha\beta}\label{deff} \ee and of the
``antiselfdual exterior 2-forms''
$$
\qquad\qquad
f'{}^{\alpha'\beta'}:=\P_s{}^{\alpha'\beta'}_{\gamma'\delta'}
\epsilon_{\gamma\delta}\xi^{\gamma\gamma'}\xi^{\delta\delta'}
=\epsilon_{\alpha\beta}\xi^{\alpha\alpha'}\xi^{\beta\beta'}
=(\xi^T\epsilon\xi)^{\alpha'\beta'}.  \quad \eqno{(\ref{deff})'}
$$
Instead of $f^{\alpha\beta}$ (resp. $f'{}^{\alpha'\beta'}$) we shall
also adopt the matrix elements of
$\xi\bar\xi$ (resp. $\bar\xi\xi$), because \be
(\xi\bar\xi)^{\alpha\beta}=f^{\alpha\gamma}\epsilon^{\gamma\beta},
\qquad \qquad   (\bar\xi\xi)^{\alpha'\beta'}=
\epsilon^{\alpha'\gamma'}f'{}^{\gamma'\beta'}.    \label{deff"}
\ee
As when $q=1$, (only) three out of the four matrix elements
$f^{\alpha\beta}$ (resp. $f'{}^{\alpha'\beta'}$) are independent,
because (\ref{xixirel}) implies
$\epsilon_{\alpha\beta}f^{\alpha\beta}=0=\epsilon_{\alpha'\beta'}f'{}^{\alpha'\beta'}$.
Together, these $f^{\alpha\beta},f'{}^{\alpha'\beta'}$ form a basis
of the 6-dimensional $\A$-bimodule (resp. ${\cal H}$-bimodule)
 $\Omega^2$ (resp. ${\cal DC}^2$).
Using relations (\ref{xixirel}) and (\ref{Rexpl}) one easily derives
the  following relations \bea &&x^{\alpha\alpha'}f^{\beta\gamma}=
q(\hat R_{12}\hat R_{23})^{\alpha\beta\gamma}_{\lambda\mu\nu}
f^{\lambda\mu}x^{\nu\alpha'},                    \label{xfrel}\\
&&\partial^{\alpha\alpha'}f^{\beta\gamma}=q^{-1}(\hat R_{12}\hat
R_{23})^{\alpha\beta\gamma}_{\lambda\mu\nu}
f^{\lambda\mu}\partial^{\nu\alpha'}.                \label{dfrel}
\eea The second is obtained from the first and Remark 1. In
(\ref{deff})$'$ and in the sequel we label a
formula regarding antiselfdual 2-forms
adding a prime  to the label of its selfdual
counterpart and omit it, when it can be obtained from the former by the
obvious replacements. As another example,
$$
x^{\alpha\alpha'}f'{}^{\beta'\gamma'}=
q(\hat R_{12}\hat R_{23})^{\alpha'\beta'\gamma'}_{\lambda'\mu'\nu'}
f'{}^{\lambda'\mu'}x^{\alpha\nu'}.              \eqno{(\ref{xfrel})'}
$$
From the previous three formulae and (\ref{dfrel})$'$ it follows
that $\Omega^2$ (resp. ${\cal DC}^2$) splits into the direct sum of
$\A$- (resp. ${\cal H}$-) bimodules \be
\Omega^2=\check\Omega^2\oplus \check\Omega^{2}{}' \qquad\quad
\mbox{(resp. }{\cal DC}^2=\check{\cal DC}^2\oplus \check{\cal
DC}^{2}{}'\mbox{)}                             \label{split} \ee of
the  eigenspaces of $*$ with eigenvalues $1,-1$ respectively.
In \cite{Fio06} we prove that
\be
\omega_2\,\omega_2'=\omega_2'\,\omega_2=0       \label{ASortho}
\ee
for any $\omega_2\in\check\Omega^2$.
$\omega_2'\in\check\Omega^2{}'$,  (resp. $\omega_2\in\check{\cal
DC}^2$, $\omega_2'\in\check{\cal DC}^2{}'$)



The 2-forms 
$(\xi\bar\xi)^{\alpha\beta}$, $(\bar\xi\xi)^{\alpha'\beta'}$
are exact.
1-form-valued matrices $a,a'$ such that
\be d\,a=\xi\bar\xi, \qquad \qquad d\,a'{}=\bar\xi\xi
\label{da=xibarxi} \ee  are clearly defined up to $d$-exact
terms. One can choose \be a_{\kappa}:=-\xi\bar x+\kappa\,\theta
|x|^2 \,I_2 \ee with complex $\kappa$. If $\kappa\neq
\kappa_0:=q^2(q^2\!-\!1)/(q^2\!+\!1)$ the four matrix elements of
$a_{\kappa}$ are all independent and make up an alternative basis
for $\Omega^1$; they belong to the $(3,1)\oplus (1,1)$-dimensional
(reducible) corepresentation of $SU_q(2)\times SU_q'(2)$. (And
similarly for $\hat a'_{\kappa}$). Whereas there are only three
independent \be a_{\kappa_0}{}^{\alpha\beta}={\cal
P}_s{}^{\alpha\lambda}_{\gamma\delta} (\xi\epsilon
x^T)^{\gamma\delta} \epsilon^{\beta\delta}, \label{aexplicit} \ee
because
$a_{\kappa_0}{}^{\alpha\beta}(\epsilon\epsilon^T)_{\beta\alpha}=0$;
the latter belong to the (3,1) irreducible corepresentation of
$SU_q(2)\times SU_q'(2)$.  In the $q=1$ limit (\ref{aexplicit})
becomes the familiar
\[
a_{\kappa_0}{}^{\alpha\beta}=-\Big(\xi\epsilon^{-1}x^T\Big)^{(\alpha\lambda)}
\epsilon^{\lambda\beta}=-\left\{Im(\xi\,\bar
x)\right\}^{\alpha\beta},
\]
where $(\alpha\lambda)$ denotes symmetrization w.r.t. $\alpha,\lambda$, and $Im$ the
imaginary part. Another peculiar choice is $\kappa_+=q\!-\!1$, which
gives $a_{\kappa_+}=-q(dT)\overline{T}|x|^2$, whence the simple change
\be \overline{T}[(dT)\overline{T}] T=\overline{T}(dT) \ee under the
`similarity' transformation $T$,  as when $q=1$. Since
$(\bar\xi\xi)^{\alpha'\beta'}$ belongs to $\check\Omega^2{}'$, which is a
$\A$-bimodule, we also find [using (\ref{theta|x|rel}) and (\ref{xblu})]
\be T\bar\xi\xi\overline{T}  =\xi\bar\xi q^2
+(q^{-2}\!-\!q^2)\xi\bar x\theta \:\in\check\Omega^2{}' .
\label{salabim} \ee

\section{Formulations of NC gauge theories}
\label{gaugeth}

We recall some minimal common elements in the formulations of
$U(n)$ gauge theories
on commutative as well as noncommutative spaces \cite{Con94,Mad99}
 (see also \cite{Lan97,FigGraVar01}).
We denote by $\A$ the `$\star$-algebra of functions on the noncommutative
space' under consideration, by $(d,\Omega^*)$ a differential calculus
on $\A$, real in the sense that  $d(f^{\star})= (df)^{\star}$.
 In $U(n)$ gauge
theory the gauge transformations $U$ are unitary $\A$-valued
$n\times n$  matrices, 
$U\!\in\! M_n(\A)\equiv M_n(\b{C})\otimes_{\b{C}}\A$ with 
$U^\dagger=U^{-1}$. The gauge potential $A\equiv (
A^{\dot\alpha}_{\dot\beta})$ is an antihermitean 1-form-valued
$n\times n$ matrix, $A\in M_n(\Omega^1(\A))$ with $A^\dagger=-A$.
The associated field strength $F\in M_n(\Omega^2)$ and
covariant derivative $D:M_n(\Omega^p)\to M_n(\Omega^{p+1})$
are defined as usual by
\be
F:=dA+AA\qquad\qquad D\omega_p:=d\omega_p+[A,\omega_p\},
\ee
and are therefore hermitean,
in the sense $F^{\dagger}=F$, $D(f^{\dagger})= (Df)^{\dagger}$.
At the right-hand side the product $AA$ has to be
understood both as a (row by column) matrix product and as a wedge
product. Even for $n=1$ (electromagnetism) $AA\neq 0$,
contrary to the commutative
case. The Bianchi identity $DF=dF+[A,F]=0$ is automatically
satisfied and the Yang-Mills equation reads as usual $D{}^*F=0$.
Because of the Bianchi identity, in a 4D Riemannian
geometry endowed with a Hodge map $*$ the latter is automatically
satisfied by any solution of the (anti)self-duality equations \be
{}^*F=\pm F. \label{ASDEq} \ee
If $\Omega^2$ splits as in (\ref{split}) then $F$ is uniquely decomposed in
a selfdual and an antiselfdual part,
\be
F=F^++F^-.                      \label{Fsplit}
\ee

The Bianchi identity, the  Yang-Mills equation, the
(anti)self-duality equations,  the
flatness condition $F=0$ are preserved by gauge transformations
\be
A^U=U^{-1}(AU +dU), \qquad\Rightarrow \qquad F^U= U^{-1}F U.  \label{gaugetransf}
\ee
As usual, $A=U^{-1}dU$ implies $F=0$.
If the exterior derivative can be
realized as the graded commutator (\ref{thetacommu}) with a special
1-form \cite{Con94,Wor89,Mad99} $-\theta$, then introducing the
1-form-valued matrix $B:=-\theta I_n+A$ one finds that
\be
F=BB,
\qquad \qquad D=[B,\cdot~\}
\ee
and Bianchi identity is now even more
trivial. In Connes' noncommutative geometry $-\theta$ is the
'Dirac operator', which has to fulfill more stringent requirements
\cite{Con94}.

Up to normalization factors,
the gauge invariant `action' $S$ and `Pontryagin index'
(or `second Chern number') $\c{Q}$
are defined by
\be
S = \mbox{Tr}(F\:{}^*\!F), \qquad\qquad\qquad  \c{Q} = \mbox{Tr}(FF)
              \label{actionfun}
\ee
where Tr stands for  a
positive-definite trace combining the $n\times n$-matrix trace with
the integral over the noncommutative manifold (as such, Tr has to
fulfill the cyclic property). If integration $\int$ fulfills itself
the cyclic property then this is obtained by simply choosing $
\mbox{Tr}=\int  \mbox{tr}$, where $\mbox{tr}$ stands for the
ordinary matrix trace.
If, as in the case under discussion,  (\ref{ASortho}) holds,
$S,\c{Q}$ respectively split into the sum, difference of two nonnegative
contributions:
\be
S =
\mbox{Tr}(F^+{}^*\!F^+)\!+\!\mbox{Tr}(F^-{}^*\!F^-),
\qquad\qquad\c{Q}  = \mbox{Tr}(F^+\:{}^*\!F^+)-\mbox{Tr}(F^-\:{}^*\!F^-).
\ee
As in the commutative case, these relations imply
$S\ge |\c{Q}|\ge 0$.

In commutative geometry the socalled Serre-Swan theorem \cite{Ser62,Con95}
states that vector bundles over a compact manifold coincide
with finitely generated projective modules
${\cal E}$ over $\A$. The gauge connection $A$ of a gauge group
(fiber bundle) acting on a vector bundle is expressed in terms
of the projector ${\cal P}$ characterizing the projective module. Therefore
these projectors can be used to  completely determine the connections.
In Connes' standard approach \cite{Con94} to noncommutative geometry
the finitely generated projective modules are the primary objects to
define and develop the gauge theory. The topological
properties of the connections can be classified in terms of topological
invariants (Chern numbers), and the latter can be computed
directly in terms of characters of  ${\cal P}$ (Chern-Connes
characters), in particular $\c{Q}$ can be computed in terms of
the second Chern-Connes character, when
Connes' formulation of noncommutative geometry applies.

\bigskip

In the present $\A\equiv C(\b{R}_q^4)=C(\b{H}_q)$ case there are {\bf 2 main
problems} preventing the application of this formulation of gauge theories:

\begin{enumerate}

\item Integration over $\b{R}_q^4$ fulfills a {\it deformed} cyclic property
\cite{Ste96}.

\item $d(f^{\star})\neq (df)^{\star}$, and there is no
$\star$-structure $\star:\Omega^*\to\Omega^*$, but (as mentioned
in section \ref{covdiffcal}) only a
$\star$-structure  $\star:{\cal DC}^*\to {\cal DC}^*$ \cite{OgiZum92},
with a nonlinear character.

\end{enumerate}

On the basis of our results \cite{Fio04}
we hope that both problems could be solved
\begin{enumerate}

\item allowing for ${\cal DC}^1$-valued $A$ ($\Rightarrow$
${\cal DC}^2$-valued $F$'s), and/or

\item realizing Tr$(\cdot)$
by in the form $\mbox{Tr}(\cdot)$:=$\int\mbox{tr}(W\cdot)$, with $W$ some
suitable positive definite ${\cal H}$-valued (i.e.
pseudo-differential-operator-valued) $n\times n$ matrix (this
implies a change in the hermitean conjugation of differential
operators), or even a more general form.

\end{enumerate}

\section{$q$-deformed $su(2)$ instanton}
\label{instconst}

We look for  $A\in M_2(\Omega^1)$ solutions of the
(anti)self-duality equations (\ref{ASDEq}) virtually yielding a finite action
functional (\ref{actionfun}). Among them we expect
deformations of the (multi)instanton solutions of $su(2)$ Yang-Mills theory on
the ``commutative'' $\b{R}^4$. We first recall the instanton solution
of Belavin {\it et al.}
\cite{BelPolSchTyu75}, which we write down both in t' Hooft
\cite{tHo76} and in ADHM \cite{AtiDriHitMan78}  quaternion notation:
\bea
A &=&  dx^i\, \sigma^a\,
\underbrace{\eta^a_{ij}x^j \frac 1{\rho^2+r^2/2}}_{A^a_i}
= -Im\left\{\xi\,\frac{\bar x}{|x|^2}\right\} \frac
1{1+{\rho^2}\frac 1{|x|^2}}\nn  &=& -(dT)\overline{T} \frac
1{1+{\rho^2}\frac 1{|x|^2}}, \label{Belinst}\\
 F &=& \xi\bar\xi\,\rho^2\frac
1{(|x|^2+\rho^2)^2}. \nonumber
\eea Here $r^2:=x\cdot
x=2|x|^2$, $\sigma^a$ are the Pauli
matrices, $\eta^a_{ij}$ are the so-called t' Hooft $\eta$-symbols,
$\rho$ is the size of the instanton (here centered at the origin).
The third equality is based on the identity
$$
\xi\,\frac{\bar x}{|x|^2}=(dT)\overline{T}+I_2\frac {d|x|^2}{2|x|^2}
$$
and the observation that the first and second term at the rhs
 are respectively antihermitean and hermitean, i.e. the
imaginary and the real part of the quaternion.

\medskip
In terms of the modified gauge potential  $B:=A-\theta I_2$ a
natural Ansatz for the deformed instanton solution in the `regular
gauge' is (in matrix notation)
\be
B:=A-\theta I_2=\xi\frac {\bar x}{|x|^2}\,l+\theta\, I_2\,m,    \label{BAnsatz}
\ee
where $l,m$ are functions of $x$ only through $|x|$. For any $f(x)$ we shall
denote $f_q(x):= f(qx)$. Using $\theta^2=0$ and (\ref{theta|x|rel}),
(\ref{xblu}),  (\ref{thetacommu}), (\ref{blu}) we find
$$
F = B^2=\xi\bar\xi(m\!- \!l)\,l_q\frac {q^{-2}}{|x|^2}
+\xi\theta\bar x\left[(q^2\!-\!1)l_ql\!+\!l_qm\!- \!q^2m_q l\right] \frac
{q^{-2}}{|x|^2}.
$$
A sufficient condition for $F$ to be selfdual is that
the expression in the
square bracket vanishes. Setting $h:=m/l$ this amounts
to the equation $q^2h_q\!-\!h=(q^2\!-\!1)$, which is solved by
$m=\left[1+\bar\rho^2/|x|^2\right]l$,
where $\bar\rho^2$ is a constant, or might be a further generator
of the algebra, commuting with $\theta$.
Replacing in the expression
for $A,F$, we find a family of solutions
\be
\ba{l}
A_l =q(dT)\overline{T} l+ \theta I_2\,
\left\{1\!+\!\left[q\!+\!\bar\rho^2\frac 1{|x|^2}\right]l\right\},
\qquad\qquad
F_l = \xi\bar\xi\frac 1{|x|^2}\bar\rho^2 \frac {q^{-2}}{|x|^2}l_ql,
\ea
\label{1ista}
\ee
parametrized by the function $l(|x|)$. This large (compared to the
undeformed case) freedom in the choice of the solution is due
to the fact that we have not yet imposed on $A$ the antihermiticity
condition. Actually, we don't know yet what the `right' antihermiticity
condition is: in fact, for no $l$ is $A$ antihermitean
w.r.t. the $\star$-structure \cite{OgiZum92} mentioned
in section \ref{covdiffcal}. In any case, one should check that for the
final $A$ the
resulting $F$ decreases faster than $|x|^{-2}$ at infinity, so that
the resulting action functional (\ref{actionfun}) is finite.

The second term in (\ref{1ista})$_1$
is proportional to $d |x|^2$; in the commutative limit $q=1$ it is a
connection associated to the noncompact factor $GL^+(1)$ of $\b{H}$.
In this limit the
antihermiticity condition on $A$ amounts to the vanishing
of this term and
completely determines the solution. It factors $GL^+(1)$ out
of the gauge group to leave a pure $su(2)$ gauge theory.
In the $q$-deformed case, as we still ignore what the `right'
$\star$- (i.e. Hermitean) structure could be, it could well happen that
w.r.t. the latter the  second term in (\ref{1ista})$_1$ contains also a
antihermitean (i.e. imaginary) part, which would be the connection associated
to an additional $U(1)$ factor of the gauge group and which could
not be consistently disposed of. In the latter case
the associated gauge theory would necessarily be a deformed
$u(2)$ one.

For the moment we cannot solve the ambiguity, and content ourselves
with writing the solution for a couple of selected choices of $l$.
If we choose $l$ so that the second term in (\ref{1ista})$_1$ vanishes
and set $\rho^2=\bar \rho^2q^{-1}$ we obtain
\be
\ba{l}
A=-(dT)\overline{T} \frac 1{1+\rho^2\frac 1{|x|^2}}
\qquad\qquad 
F = q^{-1}\xi\bar\xi\frac 1{q^2|x|^2+\rho^2} \rho^2
\frac 1{|x|^2+\rho^2}.
\ea
\label{lista1}
\ee
This has manifestly the desired $q\to 1$ limit (\ref{Belinst}). The second
choice,
$$
\ba{l}
l=-\frac{1\!+\!q^2}{1\!+\!q^4}\frac
1{1\!+\!\tilde\rho^2\frac 1{|x|^2}}\qquad\qquad
\tilde\rho^2:=\frac{1\!+\!q^2}{1\!+\!q^4}\bar\rho^2,
\ea
$$
is designed in order that $A$ is proportional to the $a_{\kappa_0}$
of (\ref{aexplicit}), so that $A^{\alpha\beta}$
span the (3,1) dimensional, irreducible corepresentation of
$SU_q(2)\times SU_q'(2)$. The result is:
\be
\ba{l}
\tilde A=-\frac{1\!+\!q^2}{1\!+\!q^4}a_{\kappa_0}\frac
1{|x|^2+\tilde\rho^2}\qquad\qquad 
\tilde F =\frac{1\!+\!q^2}{1\!+\!q^4}\xi\bar\xi\,\frac 1{q^2|x|^2+
\tilde\rho^2} \tilde\rho^2 \frac 1{|x|^2+\tilde\rho^2}. \ea
\label{1ista2}
\ee
This also  has the desired $q\to 1$ limit (\ref{Belinst}).
If $\bar \rho^2\neq 0$, in both cases $FF$ is regular everywhere and
decreases as $1/|x|^8$ as $x\to \infty$, therefore it virtually will yield
finite action $S$ and Pontryagin index  $\c{Q}$ upon integration.

As in the undeformed case, to make the determination of
multi-instanton solutions easier it is useful to go
to the ``singular gauge''. Note that as in the $q=1$
case $T=x/|x|$ is unitary and formally not continuous at $x=0$, so it can play
the role of a `singular gauge transformation'. In fact $A$ can be
obtained through the gauge transformation $A=T(\hat
A\overline{T}+d\overline{T})$ from the ``singular'' gauge potential
\bea
\hat A &=& \overline{T}dT\frac 1{1+|x|^2\frac 1{\rho^2}} \label{hatA0}\\
&\stackrel{(\ref{dT})}{=}&- \left[q^{-1}\bar\xi\frac x{|x|^2}-\frac
{q^{-3}}{q\!+\!1}
\xi^{\alpha\alpha'}\frac{x^{\beta\beta'}}{|x|^2}\epsilon_{\alpha\beta}
\epsilon_{\alpha'\beta'}\right]\frac 1{1+|x|^2\frac 1{\rho^2}}  \label{hatA}\\
\hat F &=& \overline{T}q^{-1}\xi\bar\xi\frac 1{q^2|x|^2+\rho^2}
\rho^2 \frac 1{|x|^2+\rho^2} T, \eea which is the analog of the
instanton  solution in the ``singular gauge'' found by 't Hooft in
\cite{tHo76}. $\hat A$ is singular in that it has a pole in $|x|=0$.
More generally, the generic solution (\ref{1ista}) can be obtained
through the gauge transformation $A_l=T(\hat
A_l\overline{T}+d\overline{T})$ from a singular solution $\hat A_l$.
The latter can be obtained also by starting from an Ansatz like
$\hat B=\bar\xi\frac { x}{|x|^2}\,\hat l+\theta\, I_2\,\hat m$,
instead of (\ref{BAnsatz}), and imposing that the $\bar\xi\xi$ and
the $\bar\xi\theta x$ term in $\hat F=\hat B^2$ appear in a
combination proportional to (\ref{salabim}).

A straightforward computation by means of (\ref{utile4}) shows that
$\hat A$ can be expressed also in the
form
\be
\hat A= \big(\hat{\cal D}\phi\big)\phi^{-1},              \label{Dphi}
\ee
where $\hat{\cal D}$ is the first-order-differential-operator-valued
$2\times 2$ matrix obtained from the
expression in the square bracket in (\ref{hatA}) by the
replacement $x^{\alpha\alpha'}/|x|^2\to q^4\partial^{\alpha\alpha'}$,
\be
\hat{\cal D}:=q^3\bar\xi\partial-\frac {q}{q\!+\!1}d I_2, \label{Doper}
\ee
(for simplicity we are here assuming that $\rho^2$ commutes with
$\xi\partial$) and $ \phi$ is the harmonic potential
$$
\phi:=1+\rho^2\frac 1{|x|^2}, \qquad\qquad \Box\phi=0.
$$
This is the analog of what happens in the classical case.

\bigskip

The {\bf anti-instanton solution} is obtained just by converting
unbarred into barred matrices, and conversely, as in the $q=1$ case.
For instance, from (\ref{lista1}) we obtain the anti-instanton
solution in the regular gauge \be \ba{l} A' = - (d\overline{T})T
\,\frac 1{1+{\rho^2}\frac
1{|x|^2}},\qquad\qquad 
F' = q^{-1}\bar\xi\xi \frac 1{|x|^2+\rho^2}
\rho^2\frac 1{q^2|x|^2+\rho^2},\ea\label{qantiinst}
\ee
and for the one in the singular gauge
$\hat A'= \big(\hat{\cal D}'\phi\big)\phi^{-1}$, where \be
\hat{\cal D}':=q^3\xi\bar\partial-\frac {q}{q\!+\!1}d I_2. \label{Doper'}
\ee

\subsubsection*{Recovering the instanton projective module of
\cite{DabLanMas01}} \label{compare}

In commutative geometry
the instanton projective module ${\cal E}$ over $\A$ and the associated gauge
connection can be most easily obtained using the quaternion formalism, in the
way described e.g. in Ref. \cite{Ati79}. $\b{H}\sim\b{R}^4$
can be compactified as $P^1(\b{H})\sim S^4$. Let $(w,x)\in\b{H}^2$ be
homogenous coordinates of the latter, and choose $w=I_2$ on the chart
$\b{H}\sim\b{R}^4$. The element  $u\in\b{H}^2$ defined by
\be
u \equiv \left(\ba{c} u_1\\ u_2\ea \right)=
\left(\ba{c} I_2\\ \frac{\rho x}{|x|^2}\ea \right)
\left(1\!+\!\frac{\rho^2}{|x|^2}\right)^{-1/2}   \label{defu}
\ee
fulfills $u^{\dagger}u=I_2\1$, and the $4\times 2$ $\A$-valued  matrix
$u$ has only three independent components.
Therefore the $4\times 4$ $\A$-valued matrix
\be
{\cal P}:=uu^{\dagger}=\left(\ba{lll} I_2 & \frac{\rho \bar x}{|x|^2}\\
\frac{\rho x}{|x|^2} & \frac{\rho^2}{|x|^2}I_2 \ea \right)
\frac 1{1\!+\!\frac{\rho^2}{|x|^2}}                    \label{calP}
\ee
is a self-adjoint three-dimensional projector. It is
the projector associated in the Serre-Swan theorem
correspondence to the gauge connection (\ref{hatA0}), by the formula $\hat
A=u^{\dagger}du$.
The associated projective module ${\cal E}$ is embedded
in the free module $\A^{16}$ seen as $M_4(\A)$, and is obtained from the latter
as ${\cal E}={\cal P}M_4(\A)$.

In the present $q$-deformed setting we immediately check that
the element $u\in\b{H}_q^2$ defined by (\ref{defu}) fulfills
$u^{\dagger}u=I_2\1$ again, so that
the $4\times 2$ $\A$-valued  matrix ${\cal P}$ defined by (\ref{calP})
is hermitean and idempotent, and has only 3 independent components.
Therefore, it defines the `instanton projective module'
${\cal E}={\cal P}M_4(\A)$ also in the $q$-deformed case.
One can easily verify that ${\cal P}$ reduces to the hermitean idempotent
$e$ of \cite{DabLanMas01} if one chooses the instanton
size as $\rho=1/\sqrt{2}$ and performs the change of generators
(\ref{redef}). Therefore, interpreting the model  \cite{DabLanMas01}
as a compactification to $S_q^4$ of ours, we can use all
the results \cite{DabLanMas01} about the Chern-Connes classes of $e$.

Unfortunately in the  $q$-deformed case it is no more true that
$\hat A=u^{\dagger}du$,
essentially because the $|x|$-dependent global factor multiplying
the matrix at the rhs(\ref{calP}) does not commute with the 1-forms
of the present calculus ($|x|\xi^i=q\xi^i|x|$).

\section{Changing size and center of the (anti)instanton}
\label{shift}

Applying the $\widetilde{SO_q(4)}$ coaction
(\ref{SUq2SUq2coaction}) to  the instanton  gauge potentials (\ref{1ista})
we find
\be
A_l(\xi,x)\stackrel{\Delta^{(2)}}{\longrightarrow}
aA_l\big(\xi|c|,x|c|\big)a^{-1},\qquad\qquad\quad
F_l(\xi,x)\stackrel{\Delta^{(2)}}{\longrightarrow} aF_l\big(\xi|c|,x|c|\big)a^{-1}.
\ee
where $|c|^2:=|a|^2|b|^2$. The result is the same also if we
consider  $|c|^2$ as an independent parameter
and choose $a,b$ with $|a|\!=\!|b|\!=\!\1$.
In particular, on (\ref{lista1})
\be
A\stackrel{\Delta^{(2)}}{\longrightarrow}-a\,(dT)\overline{T}\frac
1{1\!+\!\rho'{}^2\frac 1{|x|^2}}a^{-1}, \qquad\qquad F\stackrel{\Delta^{(2)}}{\longrightarrow}
a\,\xi\bar\xi\frac {q^{-1}\rho'{}^2}{q^2|x|^2\!+\! \rho'{}^2}
\frac 1{|x|^2\!+\!\rho'{}^2}a^{-1},
\ee
where we have set $\rho'{}^2:=\rho^2|c|^{-2}$.
These gauge potentials are again solutions of the self-duality equation, since
the latter is covariant under the $\widetilde{SO_q(4)}$ coaction.
The result of the $SO_q(4)$ coaction ($|a|=|b|=\1$) can be
reabsorbed into a (global) gauge transformation (\ref{gaugetransf}), with
$U=a$ (and similarly $U=\bar b$  for the anti-instanton gauge
potentials), i.e. is a gauge
equivalent solution. Note that we are thus introducing
gauge transformations depending on the additional noncommuting
parameters $a,b$.  A full  $\widetilde{SO_q(4)}$ coaction ($|c|\neq 1$) instead
involves also a change of the size of the instanton, and gives an inequivalent
solution. We can thus obtain any size starting from the instanton with unit
size.

\medskip
Having built an (anti)instanton ``centered at the origin'' with
arbitrary size one would like first to translate the latter in space
to another point $y$, then to construct $n$-instanton solutions
``centered at points $y_{\mu}$'', $\mu=1,2,\ldots,n$. The
appropriate framework is to replace tensor products $\otimes$ by
braided tensor products $\uot$ and apply the braided coaddition
\cite{Maj95} to the covectors $x$. This gives new (i.e. gauge
inequivalent) solutions. The braided coaddition \cite{Maj95} of the
coordinates $x$ reads $\und{\Delta}(x)\!=\!x\uot {\bf 1}\!+\!{\bf 1}\uot
x \!\equiv\! x\!-\!y$, where we have renamed $x:= x\uot{\bf 1}$,
$y\!:=\!-{\bf 1}\!\uot\! x$.
It follows $y\bar y=\bar yy=I_2|y|^2$. Out of the two possible braidings we choose the
following one: \bea &&  y^{\alpha\alpha'}x^{\beta\beta'}= \hat
R^{\alpha\beta}_{\gamma\delta} \hat
R^{\alpha'\beta'}_{\gamma'\delta'}
x^{\gamma\gamma'}y^{\delta\delta'},\nn && \partial_{\alpha\alpha'}
y^{\beta\beta'}\!\!=\! \hat R^{\beta\delta}_{\alpha\gamma} \hat
R^{\beta'\delta'}_{\alpha'\gamma'}
y^{\gamma\gamma'} \!\partial_{\delta\delta'}, \qquad\:    \label{yrel}\\
&&  y^{\alpha\alpha'}\xi^{\beta\beta'}= \hat
R^{\alpha\beta}_{\gamma\delta} \hat
R^{\alpha'\beta'}_{\gamma'\delta'}
\xi^{\gamma\gamma'}y^{\delta\delta'}.\nonumber \eea
We also enlarge the algebra by introducing further generators $1/|y|$,
$1/|z|$ (where $z:=x\!-\!y$) fulfilling relations
\be
\ba{l}
q\gamma\xi^i = \xi^i \gamma, \qquad \qquad \mbox{for }\qquad \gamma=
\frac 1{|x|},\frac 1{|y|}, \frac 1{|z|} \\ [8pt]
 y^i \frac 1{|x|}\!=\!\frac q{|x|}y^i, \qquad
 x^i\frac q{|y|} \!=\!\frac 1{|y|}  x^i, \qquad \frac 1{|y|}\frac
1{|x|} \!=\!\frac q{|x|}\frac 1{|y|},\\[8pt]
\frac 1{|z|} \frac{x^i}{|x|^2}=\frac {x^i}{|x|^2}\frac q{|z|}
\!+\!(1\!-\!q)\frac{z^i}{|z|^3}, \qquad \frac q{|z|}
\frac{y^i}{|y|^2}=\frac {y^i}{|y|^2}\frac 1{|z|}
\!+\!(1\!-\!q)\frac{z^i}{|z|^3}.
\ea\qquad\qquad
 \ee
These are (the only) consistent extensions of the previous relations
to the inverse square root of $|z|^2,|x|^2, |y|^2$ having the
desired, commutative $q\to 1$ limit.

Under the replacement $x\to x-y$ (i.e. under $\und{\Delta}$) the
differential calculus is invariant, implying that solutions are
mapped into solutions. Therefore the instanton solution with
``shifted'' center $y$ will read in the regular gauge \be \ba{l}
A=-(dT)\overline{T} \frac 1{1+\rho^2\frac 1{|x-y|^2}}
\qquad\qquad 
F = q^{-1}\xi\bar\xi\frac 1{q^2|x-y|^2+\rho^2} \rho^2
\frac 1{|x-y|^2+\rho^2}.
\ea
\label{1istsh}
\ee
and in the singular gauge
\be
\ba{l}
\hat A= \big(\hat{\cal D}\phi\big)\phi^{-1},
\qquad\qquad \phi:=1+\rho^2\frac 1{|x-y|^2},\\
\hat F =\overline{T}q^{-1}\xi\bar\xi T\frac 1{q^2|x-y|^2+\rho^2} \rho^2
\frac 1{|x-y|^2+\rho^2}.
\ea                                   \label{phi-Ash}
\ee

\section{Multi-instanton solutions}
\label{multiinst}

On the basis of the latter and of the $q=1$  results
\cite{tHo76,vari}, we first look for $n$-instanton solutions of the
self-duality equation in the ``singular gauge'' in the form
(\ref{Dphi}). Beside the coordinates $x^i\equiv -y_0^i$ we introduce
$n$ other coordinates $y^i_{\mu}$, $\mu=1,2,...,n$ generating as
many $\b{R}_q^4$ and braided to each other: \be \ba{l}
y_{\mu}\bar y_{\mu}=\bar y_{\mu}y_{\mu}=I_2|y_{\mu}|^2\\[8pt]
y_{\nu}^{\alpha\alpha'}y_{\mu}^{\beta\beta'}= \hat
R^{\alpha\beta}_{\gamma\delta} \hat
R^{\alpha'\beta'}_{\gamma'\delta'}y_{\mu}^{\gamma\gamma'}y_{\nu}^{\delta\delta'}
\ea \label{yyrel} \ee with $\mu<\nu$ and no sum over repeated $\mu$. We shall 
call $\A_n$ the
larger algebra  generated by the $y_{\mu}^i$'s and by  parameters
$\rho_{\mu}$, $\mu=1,...,n$ fulfilling the commutation relations
\be
\ba{l}
\rho_{\nu}^2\rho_{\mu}^2=q^2\,\rho_{\mu}^2\rho_{\nu}^2,\qquad \qquad
\nu<\mu,           \\[8pt]
\rho_{\nu}^2y_{\mu}^i=y_{\mu}^i\rho_{\nu}^2  \cases{q^{-2}\:\quad
\nu<\mu, \cr 1, \:\quad\nu \ge\mu .}\ea\label{new3} \ee We shall also
enlarge $\A_n$ to the extended Heisenberg algebra ${\cal H}_n$ and
extended algebra of differential forms  $\Omega^*(\A_n)$ by adding
as generators the $\partial_i$ and the $\xi^i$ respectively, and to
the  extended differential calculus algebra ${\cal DC}(\A_n)$ by
adding as generators both the $\xi^i,\partial_i$, with cross
commutation relations \be
\ba{ll}
\rho_{\mu}^2\xi^{\alpha\alpha'}=\xi^{\alpha\alpha'}\rho_{\mu}^2,
\qquad\qquad\qquad
&\partial_{\alpha\alpha'}\rho_{\mu}^2=\rho_{\mu}^2\partial_{\alpha\alpha'}, \\[8pt]
y_{\mu}^{\alpha\alpha'}\xi^{\beta\beta'}= \hat
R^{\alpha\beta}_{\gamma\delta} \hat
R^{\alpha'\beta'}_{\gamma'\delta'}
\xi^{\gamma\gamma'}y_{\mu}^{\delta\delta'}, \qquad\qquad
&\partial_{\alpha\alpha'} y_{\mu}^{\beta\beta'}\!\!=\! \hat
R^{\beta\delta}_{\alpha\gamma} \hat
R^{\beta'\delta'}_{\alpha'\gamma'} y_{\mu}^{\gamma\gamma'}
\!\partial_{\delta\delta'}, \ea \ee Note that the first  relations,
together with the decomposition $d=\xi^i\partial_i$, imply \be
d\,\rho_{\mu}^2=\rho_{\mu}^2 d. \ee Also, from these relations it is
evident that $\check\Omega^2(\A_n),\check\Omega^{2}{}'(\A_n)$ are
$\A_n$-bimodules (resp. $\check{\cal DC}^2(\A_n), \check{\cal
DC}^{2}{}'(\A_n)$ are ${\cal H}_n$-bimodules). Let us introduce the
short-hand notation
$$
z_{\mu}^{\alpha\alpha'}:=x^{\alpha\alpha'}-v^{\alpha\alpha'}_{\mu},\qquad\quad
v^{\alpha\alpha'}_{\mu}:=
\sum\limits_{\nu=1}^{\mu}y^{\alpha\alpha'}_{\nu}, \qquad\qquad
\mu=1,2,...,n;
$$
$v^{\alpha\alpha'}_{\mu}$ will play the role of coordinates of the
center of the $\mu$-th instanton. It is easy to check from
(\ref{yyrel}) that these new $n$ sets of variables generate as many
copies of the quantum Euclidean space $\b{R}_q^4$, namely \be
 z_{\mu} \bar z_{\mu}= \bar z_{\mu}z_{\mu}=|z_{\mu}|^2I_2
\label{zzrel} \ee (no sum over repeated $\mu$) and together with $x^{\alpha\alpha'}$ make up an
alternative Poincar\'e-Birkhoff-Witt basis of the algebra $\A_n$,
(i.e. ordered monomials in these variables make up a basis of the
vector space underlying $\A_n$). Moreover, differentiating
$z_{\mu}^{\alpha\alpha'}$ and commuting it with $\xi^{\beta\beta'}$
is like differentiating and commuting $x^{\alpha\alpha'}$:
$$
\quad \partial_{\alpha\alpha'} z_{\mu}^{\beta\beta'}\!\!=\!\delta^{\beta}_{\alpha}
\delta^{\beta'}_{\alpha'}\!+\! \hat R^{\beta\delta}_{\alpha\gamma}
\hat R^{\beta'\delta'}_{\alpha'\gamma'} z_{\mu}^{\gamma\gamma'}
\!\partial_{\delta\delta'}, \eqno{(\ref{dxrel})}_{\mu}
$$
$$
 z_{\mu}^{\alpha\alpha'}\xi^{\beta\beta'}= \hat
R^{\alpha\beta}_{\gamma\delta} \hat
R^{\alpha'\beta'}_{\gamma'\delta'}
\xi^{\gamma\gamma'}z_{\mu}^{\delta\delta'}.
\eqno{(\ref{xxirel})}_{\mu}
$$
Therefore for any $\mu=1,2,...,n$ the replacement $x\to z_\mu$ in
any true relation involving $x,\partial,\xi$ will generate a new
true relation, which we shall label by adding the subscript $\mu$ to
the original one.

The solution $\phi$ searched for (\ref{Dphi}) is of the form \be
\phi\equiv\phi_n=1+\sum\limits_{\mu=1}^n\rho_{\mu}^2\frac
1{|z_{\mu}|^2}, \label{genphi} \ee namely a scalar ``function'' of
the coordinates $x^i$, of the instanton ``sizes'' $\rho_{\mu}$ and
of the ``coordinates of their centers''. For this to be allowed we
have further enlarged $\A_n,\Omega^*(\A_n),{\cal H}_n,{\cal
DC}(\A_n)$ to  extended algebras
$\A^{ext}_n,\Omega^*(\A^{ext}_n){\cal H}^{ext}_n,{\cal
DC}(\A^{ext}_n)$   by adding as generators inverse square roots
$1/|z_{\mu}|$, but we also add the inverses $1/\phi_m$, together
with corresponding commutation relations (see \cite{Fio06})
consistent with the ones given so far. The basic ones can be
obtained from the relations of section \ref{shift} by the
replacements $x\to z_{\mu}$,  $\rho\to \rho_{\mu}$, $y\to
\sum_{\lambda=\mu\!+\!1}^{\nu}y_{\lambda}$, $z\to z_{\nu}$,
$\rho_z\to \rho_{\nu}$ with $\nu>\mu$. By
relations (\ref{utile4}), (\ref{dxrel})$_{\mu}$ $\phi$ is harmonic,
exactly as in the classical case.   In Theorem 1 of \cite{Fio06} we
prove that $\hat A= \big(\hat{\cal D}\phi\big)\phi^{-1}$ fulfills
the selfduality equation (\ref{ASDEq})$_1$. Explicitly, the field
strength is \be \hat F= \frac
{-q^5}{4_q}\left[\epsilon^{-1}\!(\xi\bar\xi\partial)^T\!\epsilon
\partial\phi\right]\left[q\phi^{-1}
\!\!+\!\phi^{-1}_q\right]+q^2\epsilon^{-1}\!(
\xi\bar\xi\partial\phi)^T\!\epsilon(\partial \phi)\phi^{-1}
\phi^{-1}_q, \ee where $\phi_q(\{z_i\}):=\phi(\{qz_i\})$. (This is a
selfdual  matrix because $\xi\bar\xi$ is.)

Formally, as $x\to \infty$ also $z_{\mu}\to\infty$, $\phi\to 1$, and
a simple inspection shows that $\hat A\to 0$ as $1/|x|^3$, $\hat
F\to 0$ as $1/|x|^4$, exactly as in the case $q=1$. Therefore $\hat
F\hat F$ decreases fast enough at infinity for integrals like
$\int\mbox{tr}(\hat F\hat F)$ to converge.

On the other hand, as $z_{\mu}\to 0$ the function $\phi$ and
therefore the gauge potential $\hat A$  are singular, i.e. formally
diverge. We don't know yet whether the singularity will cause
problems also in a proper functional-analytical treatment (this
requires analyzing representations of the algebra). If this is the
case then, as in the undeformed theory, the question arises if this
singularity is only due to the choice of a singular gauge and can be
removed by performing a suitable gauge transformation, or it really
affects the field strength. We address this issue
semi-heuristically. We shall say that an element of our algebra is:
1. analytic in $z_{\mu}$ if its power expansion has no poles in
$z_{\mu}$, i.e. does not depend on $1/|z_{\mu}|$; regular in
$z_{\mu}$ if it formally keeps finite as $z_{\mu}\to 0$, i.e. in its
power expansion the dependence on  $1/|z_{\mu}|$ occurs only through
$z_{\mu}/|z_{\mu}|$. Since such dependences might change upon
changing the order in which the variables $z_1,z_2,...,z_n$, and
possible extra variables $1/|z_1\!-\!z_2|,1/|z_1\!-\!z_3|,...$ (if
necessary), are displayed, these conditions have to be met for any
order. In the appendix of \cite{Fio06} we show that performing the
``singular gauge transformation'' $U_2$ defined by \be U_2\equiv
U_2(z_1,z_2):=\frac{\bar z_1}{|z_1|}\frac{y_2}{|y_2|} \frac{\bar
z_2}{|z_2|}                      \label{defU_2} \ee on $\hat A_2$ we
obtain a $2$-istanton solution \be A_2=U_2^{-1}\left(\hat A
U_2+dU_2\right)         \label{A_2} \ee analytic in both $z_1,z_2$;
the corresponding  selfdual field strength will be analytic as well.
The form of $U_2$ exactly mimics the undeformed one of Ref.
\cite{GiaRot77,OliSciCre79}. Of course, for this to make sense, we
have to further enlarge the algebras adding as a generator $1/|y_2|$
with consistent commutation relations; this is done in appendix A.1
of \cite{Fio06}. By generalization of the undeformed reults
\cite{GiaRot77,OliSciCre79}, we are led to the

\medskip
{\bf Conjecture.}
Performing the singular
gauge transformation $U_n$ recursively defined by $U_0=\1_2$ and
\be
U_n\equiv U_n(z_1,...,z_n):=U_{n\!-\!1}(z_1,...,z_{n\!-\!1})
U^{-1}_{n\!-\!1}(y)\frac{\bar z_n}{|z_n|},            \label{defU}
\ee
with $U_{m}(y)$ the function of $y_1,...y_m$ only
defined by $U_m(y):=U_m(z_1\!-\!z_n,...,z_{n\!-\!1}\!-\!z_n)$, we finally obtain a
regular $n$-istanton solution
\be
A\equiv A_n=U_n^{-1}\left(\hat A U_n+dU_n\right)         \label{A_n}
\ee
and a corresponding regular selfdual field strength, for any $n$.

\bigskip
Results for the {\bf $n$-antiinstanton solutions} are obtained
by the already mentioned replacements.
In particular, the singular ones  $\hat A$
are simply obtained replacing $\hat{\cal D}$ with  $\hat{\cal D}'$
in (\ref{Dphi}).



\begin{thebibliography}{99}

\bibitem{Ati79} M. F. Atiyah, {\it Geometry on Yang-Mills fields}
 {\em  Lezioni Fermiane}, Scuola Normale Superiore di Pisa
(1979).

\bibitem{AtiDriHitMan78} M. F. Atiyah, N. J. Hitchin, V. G. Drinfel'd,
Yu. I. Manin, {\it  Construction of instantons}, {\em Phys. Lett.} {\bf  65A}
(1978), 185-187.

\bibitem{AscDimMeySchWes06}
P. Aschieri, M. Dimitrijevic, F. Meyer, S. Schraml, J. Wess,
{\it  Twisted Gauge Theories}, {\em Lett. Math. Phys. } {\bf 78} (2006), 61-71

\bibitem{BelPolSchTyu75} A.\ A.\ Belavin, A.\ M.\ Polyakov, A.\ S.\
Schwarz and Yu.\ S. Tyupkin,  {\it Pseudoparticle solutions of the Yang-Mills equations}, {\em  Phys. Lett.} {\bf  59B}
(1975), 85-87.

\bibitem{BonCicTar00}
 F. Bonechi, N. Ciccoli, M. Tarlini,  {\it Noncommutative instantons
on the 4-sphere from quantum groups}, {\em  Commun. Math. Phys.} {\bf  226}
(2002),  419-432.

\bibitem{CarSchWat91}
U.~Carow-Watamura, M.~Schlieker, S.~Watamura,
{\it $SO_q(N)$ covariant differential calculus on quantum space
and quantum deformation of Schroedinger equation},
{\em Z.Phys.} {\bf C 49} (1991), 439.

\bibitem{CerFioMad01}
B. L. Cerchiai, G. Fiore, J. Madore,
{\it  Geometrical Tools for Quantum Euclidean Spaces},
{\em Commun. Math. Phys.} {\bf 217} (2001), 521-554.

\bibitem{Con94}
A.~Connes,  {\em Noncommutative Geometry}, Academic Press, 1994.

\bibitem{Con95}
A.~Connes,  {\em Non-commutative Geometry and Physics},
Les Houches, Session LVII, Elsevier Science B. V., 1994.

\bibitem{ConLan01}
A. Connes, G. Landi, {\it Noncommutative manifolds, the instanton
algebra and isospectral deformations}, {\em Commun. Math. Phys. } {\bf 221} (2001),
141-159

\bibitem{DabLanMas01}
Ludwik Dabrowski, Giovanni Landi, Tetsuya Masuda,
 {\it Instantons on the Quantum 4-Spheres $S^4_q$},
{\em Commun. Math. Phys.} {\bf 221} (2001), 161-168.

\bibitem{DimMeyMoeWes03}
 M. Dimitrijevic, F. Meyer, L. M\"oller, J. Wess,
{\it Gauge theories on the kappa-Minkowski spacetime},
 {\em Eur. Phys. J.} {\bf C36} (2004), 117-126.

\bibitem{Dri86}
V.~Drinfeld, {\it Quantum groups,} in {\em {I.C.M.} Proceedings,
{B}erkeley}, p.~798.
\newblock 1986.

\bibitem{FadResTak89}
L. D.~Faddeev, N. Y.~Reshetikhin, L. A.~Takhtadjan,
{\it Quantization of {L}ie groups and Lie algebras},
{\em Algebra i Analyz} {\bf 1} (1989), 178, translated from the
Russian in {\em Leningrad Math. J.} {\bf 1} (1990), 193.

\bibitem{FadTak86}
L.D.~Faddeev,  L. A.~Takhtadjan,
{\it Liouville model on the lattice},
{\em Lecture Notes in Physics} {\bf 246} (1989), Springer, New York, 1986,
pp. 166-179.

\bibitem{FigGraVar01}
J. M. Gracia-Bondìa, J. C. Varilly, H. Figueroa,
{\em Elements of Noncommutative Geometry}, Birkh\"auser Boston, Inc., Boston, MA, 2001.

\bibitem{Fio94} G. \ Fiore,
{\it Quantum Groups $SO_q(N),Sp_q(n)$ have q-Determinants, too},
{\em J. Phys. A: Math Gen.} {\bf 27} (1994), 3795.

\bibitem{Fio04JPA} G. \ Fiore,
{\it Quantum group covariant (anti)symmetrizers,
$\varepsilon$-tensors, vielbein, Hodge map and Laplacian},
{\em J. Phys. A: Math. Gen.}  {\bf 37} (2004), 9175-9193.

\bibitem{Fio04} G. \ Fiore,
{\it On the hermiticity of q-differential operators and forms on the
quantum Euclidean spaces $\b{R}_q^N$}, {\em Rev. Math. Phys.} {\bf
18} (2006), 79.

\bibitem{Fio05q} G. \ Fiore, {\it On $q$-quaternions},
in preparation.

\bibitem{Fio06} G. \ Fiore, {\it $q$-Quaternions and $q$-deformed
$su(2)$ instantons},
 DSF/43-2005. hep-th/0603138

\bibitem{Fio06corfu}  G. \ Fiore,
{\it $q$-Deformed quaternions and  $\mbox{\boldmath $su(2)$}$
instantons},
Preprint 06- Dip. Matematica e Applicazioni, Universit\`a di
Napoli, and DSF/17-2006. To appear in the proceedings of the
"Noncommutative Geometry in Field and String Theories",
Satellite Workshop of "CORFU Summer Institute 2005".


\bibitem{GiaRot77}
J.\ J.\ Giambiaggi, K.\ D.\ Rothe
{\it Regular $N$-istanton fields and singular gauge transformations}
{\em Nucl. Phys.} {\bf  B129} (1977), 111-124.

\bibitem{IvaLecMue04}
T. A. Ivanova, O. Lechtenfeld, H. Mueller-Ebhardt,
{\it Noncommutative Moduli for Multi-Instantons},
{\em Mod. Phys. Lett.}  {\bf A19} (2004), 2419-2430.

\bibitem{JacReb76}
R. Jackiw, C. Rebbi, {\it Conformal properties of a Yang-Mills pseudoparticle},
  {\em Phys. Rev.} {\bf D14} (1976), 517-523.

\bibitem{JurMoeSchSchWes01}
B.\ Jurco, L.\ M\"oller, S.\ Schraml, P.\ Schupp, J.\ Wess,
{\it Construction of non-Abelian gauge-theories on noncommutative
spaces}, {\em Eur. Phys. J.} {\bf C21} (2001), 383-388.  hep-th/0104153.

\bibitem{Lan97} G. Landi,
{\it An Introduction to Noncommutative Spaces and Their Geometries},
Springer-Verlag, Berlin, 1997.

\bibitem{LanvSu06} G. Landi, W. van Suijlekom,
{\it Noncommutative instantons from twisted conformal symmetries},
math.QA70601554

\bibitem{Mad99} J. Madore, {\it An introduction to noncommutative differential geometry
and its physical applications}. Second edition.
London Mathematical Society Lecture Note Series, 257.
Cambridge University Press, Cambridge, 1999.

\bibitem{Maj92} S. Majid, {\it Braided Momentum Structure of the q-Poincare
Group}, {\em J. Math. Phys.} {\bf 34} (1993) 2045-2058.

\bibitem{Maj94} S. Majid, {\it $q$-Euclidean space and quantum group wick
rotation by twisting},  {\em J. Math. Phys.} {\bf 35}  (1994),
5025-5034.

\bibitem{Maj95} For a review see for instance: S. Majid, {\it Foundations
of Quantum Groups}, Cambridge Univ. Press (1995); and
references therein.

\bibitem{NekSch98} N. Nekrasov, A. Schwarz,
{\it Instantons on noncommutative $\b{R}^4$, and (2,0)
superconformal six dimensional theory}, {\em Commun.Math.Phys.}
{\bf  198} (1998), 689-703.

\bibitem{Ogi92}
O.\ Ogievetsky
{\it Differential operators on quantum spaces for $GL_q(n)$ and $SO_q(n)$},
{\em Lett. Math. Phys.} {\bf 24} (1992), 245.

\bibitem{OgiZum92}
O.\ Ogievetsky, B.\ Zumino
{\it Reality in the Differential calculus on the $q$-Euclidean Spaces},
{\em Lett. Math. Phys.} {\bf 25} (1992), 121-130.

\bibitem{OliSciCre79} D. I. Olive,
S. Sciuto, R. J. Crewther, {\it Instantons in field theory}, {\em Riv. Nuovo
Cim.} {\bf 2} (1979), 1-117.

\bibitem{PodWor90} P. Podl\'es, S. L. Woronowicz,
{\it Quantum deformation of Lorentz group},
{\em Commun. Math. Phys.} {\bf 130} (1990), 381-431.

%
\bibitem{S2} A. Schirrmacher, {\it Remarks on the use of $R$-matrices}, in
Quantum groups and related topics (Wroc\l aw, 1991), 55--65,{\em  Math. Phys.
Stud.} {\bf  13},    Kluwer Acad. Publ., Dordrecht, 1992.\\
A. Sudbery, {\em Math. Proc. Cambridge Philos. Soc.} {\bf  114} (1993),  111-130;
{\em Phys. Lett.} {\bf  B 284} (1992),
61-65; {\em Phys. Lett. B} {\bf  291} (1992), 519.

\bibitem{SchWatZum92}
P. Schupp, P. Watts, B. Zumino, {\it Differential Geometry on Linear Quantum
Groups}, {\em Lett. Math. Phys.}, {\bf  25} (1992), 139-148.

\bibitem{SchWeiWei92}
M. Schlieker, W. Weich and R. Weixler, {\it Inhomogenehous
quantum groups}, {\em Z. Phys.}  {\bf C 53} (1992), 79-82;

\bibitem{SeiWit99} N. Seiberg, E. Witten,
{\it String Theory and Noncommutative Geometry},
{\em JHEP} {\bf 9909} (1999) 032.

\bibitem{Ser62} R. G. Swan, {\it Vector Bundles and Projective
Modules},{\em Trans. Am. Math. Soc.} {\bf 105} (1962), 264-277.

\bibitem{Ste96} H. \  Steinacker,
{\it Integration on quantum Euclidean space and sphere
in $N$ dimensions},
{\em J. Math Phys.} {\bf 37} (1996), 4738.

\bibitem{tHo76} G. \ 't Hooft, {\it Computation of the Quantum
Effect Due to a Four-dimensional Pseudoparticle},
{\em Phys. Rev.} {\bf D14} (1976), 3432-3450.

\bibitem{Wor87}
S. L.~Woronowicz, {\it Compact Matrix Pseudogroups},
{\em   Commun.\ Math.\ Phys.} {\bf 111} (1987) 613-665.

\bibitem{Wor87a}
S. L.~Woronowicz, {\it Twisted ${SU}(2)$ group, an example of a non-commutative
  differential calculus}, {\em Publ. RIMS, Kyoto Univ.} {\bf 23} (1987) 117.

\bibitem{Wor89}
S. L.~Woronowicz, {\it Differential calculus on compact matrix pseudogroups
(Quantum Groups),}
{\em   Commun.\ Math.\ Phys.} {\bf 122} (1989) 125.

\bibitem{vari}
F.  Wilczeck, in `Quark confinement and field theory', Ed. D. Stump and D.
Weingarten, John Wiley and Sons, New York (1977).
E. Corrigan, D. B. Fairlie, {\em Phys. Lett.} {\bf 67B} (1977), 69.
R. Jackiw, C. Nohl, C. Rebbi  {\em Phys. Rev.} {\bf D15} (1977), 1642.


\end{thebibliography}
\end{document}